\documentstyle[epsf,referee]{l-aa-mpa}

\begin{document}

\thesaurus{6(06.05.1; 06.09.1; 08.05.3; 08.09.3)}

\title{The age of the most nearby star}

\author{A.~Weiss \and H.~Schlattl}

\institute{Max-Planck-Institut f\"ur Astrophysik,
           Karl-Schwarzschild-Str.~1, 85748 Garching,
           Federal Republic of Germany}

\offprints{A.~Weiss; (e-mail address: weiss@mpa-garching.mpg.de)}

\date{Received; accepted}

\maketitle
\markboth{A.~Weiss \& H.~Schlattl: The age of the most nearby star
}{The age of the most nearby star} 

\begin{abstract}
We address the question how accurately stellar ages can be determined
by stellar evolution theory. We select the star with the best
observational material available - our Sun.
We determine the solar age by fitting solar evolution models to
a number of observational quantities including several obtained from
helioseismology, such as photospheric helium abundance or p-mode
frequencies.  Different cases with respect to the number of free
parameters and that of the observables to be fitted are
investigated. Age is one of the free parameters determined by the
procedure. We find that the neglect of hydrogen-helium-diffusion leads
to ages deviating by up to 100\% from the true, meteoritic solar
age. Our best models including diffusion yield ages by about 10\% too
high. The implication for general stellar age determination is that a
higher accuracy than that can not be expected, even with the most
up-to-date models. Our results also confirm that diffusion as treated
presently in solar models is slightly too effective.
\end{abstract}

\section{Introduction}
Stellar evolution theory is able to derive the age of stars or stellar
systems from a few basic observational data and therefore can provide
constraints on the evolution of galaxies and the universe
itself. Recently, it has become evident that improvements in the input
physics used for the stellar model calculations lead to considerably
reduced ages for Pop.~II stars as observed in globular clusters
(Chaboyer \& Kim 1995; Mazzitelli et al.~1995; Salaris et al.~1997),
thereby bringing into consistency globular cluster ages and cosmic
expansion ages. In this context, the question of the accuracy of
stellar age determinations arises. The errors given for globular
cluster ages reflect the observational uncertainties
only, while those of the stellar models are usually ignored. Chaboyer
(1995) and Shi (1995) have investigated the influence of various
physical assumptions on ages determined from the main sequence
turn-off and found possible variations of up to order 10\%. One of
these -- the inclusion of non-ideal effects in the equation of state
-- resulted in the reduced cluster ages. 

The influence of the model errors on the derived ages depend, of
course, on the dating method used. One therefore tries to find ways to
minimize the influence of possible sources of error or to determine
relative ages, which in general are more accurate. For a discussion of
several methods for globular clusters, see Stetson et al.~(1996). In
general, the predicted colour and/or visual brightness of the turn-off
as a function of age is the decisive quantity, although differential
age determinators using the colour or brightness difference between
the turn-off and some time-independent cluster-diagram feature are
being used to minimize the influence of systematic offsets.

A completely independent method to determine stellar ages is used for
eclipsing binary systems. One can, for example, try to match the
positions in the Hertzsprung-Russell- or Colour-Magnitude-Diagram of
both components of known mass with stellar evolution tracks. A
necessary condition for a credible age-determination is then that the
two error boxes are hit for the same age along both tracks. The
classical example for such a successful age determination is AI~Phe
(Andersen et al.~1988) with determined ages that agree better than
10\%. Schr\"oder et al.\ (1997) recently have investigated $\zeta$~Aur
systems in a different, but related context. In the future, detached
eclipsing binary systems detected in 
massive photometric searches (e.g.\ Kaluzny et al.~1996) will allow to
determine globular cluster ages much more directly (see Paczy\'nski
1996a for a review). 
For that purpose, theoretical luminosity--age--relations for
a variety of stellar masses and compositions will be necessary (Weiss
\& Schlattl 1997) that will yield the age (and possibly the helium
content; cf.\  Paczy\'nksi 1996a) of cluster stars of known mass,
metallicity and luminosity. Again, the question will be: how accurate
are such age determinations in terms of the theoretical uncertainties?

To answer this and other questions concerning the accuracy of
stellar ages, Paczy\'nksi (1996b) suggested to investigate how
accurate we can determine the age of the best-known star - our Sun. In
calculating solar models, the solar age as determined from meteorites
usually is one of the parameters to be fitted by the model. Along with
luminosity, effective temperature and the metal-to-hydrogen ratio
$Z/X$ it determines the model parameters (initial helium content;
mixing-length parameter); the solar model then provides predictions
for the neutrino fluxes, p-mode frequencies and other
quantities. Since the agreement between models and solar properties
determined from helioseismological observations is very good, this is
implicitely taken as evidence that the solar age is appropriate for the
models. However, there are indications that a slightly different age
might lead to even better agreement. Dziembowski et al.\ (1994) have
investigated a solar seismic model of only 4 billion years. Although
they exclude this model, they also state that a much smaller change in
age $(\pm 5\cdot 10^7$ years) cannot be ruled out by helioseismology. 
Schlattl et al.\ (1997) find that their solar model would agree 
better with observations, if the solar age would be raised.

In the present paper, we will turn around the standard solar model
approach. A set of observables will be given which the models have to
match. The
solar age will now be a quantity resulting from the best fit to the
observables. That way, we will be able to determine the solar age from
stellar models. The
comparison with the meteoritic ``true'' age will give us indications
how accurate stellar ages can be determined for the case of the most
precisely known stellar parameters. For any other star, the accuracy
will be lower. It will also define the best set of input physics
assumptions able to yield the most accurate ages. Since the solar mass
and luminosity are among the known parameters, the solar case is a
stringent test case for age determinations of globular clusters based
on detached eclipsing binaries.

In the next section we will shortly review the numerical code and the
input physics assumptions of our solar model calculations published in
Schlattl et al.\ (1997). Section~3 will present solar age
determinations based on fits to different sets of solar quantities,
among them interior sound speed and p-mode frequencies. The final
section will contain our conclusions. 

\clearpage

\section{The solar model calculations}
For the present work, we have used the same solar model code as
described in Schlattl et al.\ (1997). Here, we merely recall its basic
features and input physics details.

The opacities are combinations of the OPAL-tables (Rogers \& Iglesias
1992; Iglesias \& Rogers 1996) with either those of Weiss et
al.\ (1990) or Alexander \& Fergusson (1994) for the low-temperature
regime. For the EOS we 
either use a Saha-type equation or the OPAL equation of state (Rogers
et al.\ 1996). If diffusion is included, the diffusion coefficients are
calculated according to Thoul et al.\ (1994). For this work, only
hydrogen-helium-diffusion has been taken into account. For the layers
above $\tau\approx 20$ model atmospheres have been used for the
calculations of the p-mode frequencies and as the outer boundary
condition for the inner model. Since the model atmospheres extend into
the convective envelope and take into account convection according to
the mixing-length description with a parameter $\alpha_{\rm at}=0.5$,
no constant value for the mixing-length parameter $\alpha_{\rm int}$
can be used. Rather, a smooth transition has been achieved by using the
function
\begin{eqnarray}
\alpha(T) & = & f(T) \times \alpha_{\rm int} + \left( 1-f(T)\right)
\times \alpha_{\rm at} \nonumber\\ 
\rm where \nonumber\\ 
f(T) & = & \left( 1 +
\exp\left(\frac{T_0-T}{\Delta T}\right) \right)^{-1}
\label{e:alpha}
\end{eqnarray}
and the parameters $T_0$ and $\Delta T$ being 10\,800 resp. 180 K. 
$\alpha_{\rm int}$ remains the parameter to be determined by
the solar model and replaces the global value $\alpha_{\rm MLT}$ used in
models with standard grey atmospheres. The calculations of Sect.~3.1 were
performed with those standard atmospheres and a constant $\alpha_{\rm
MLT}$. 

The composition changes due to nuclear reactions are calculated using
a network incorporating the p-p and CNO-cycle using the same reaction
rates as in Castellani et al.\ (1994). The calculations are started on
the pre-main sequence with a homogeneous model powered by gravothermal
energy only.
Both the spatial and temporal resolution are checked for their accuracy
and therefore allow a controlled precision of the models. For further
details the reader is refered to Schlattl et al.\ (1997).

Table~\ref{solval} lists the solar values adopted and the references for the
quantities resulting from helioseismological observations and
inversions. 

\begin{table}
\caption{Solar values adopted in this paper and references for
quantities inferred from helioseismology. $_\odot$ denotes 
the present Sun. $Y_\odot$ is the present surface helium content and 
$R_{\rm cz}$ the radius of the convective envelope's lower boundary.
$\nu$ indicates symbolically p-mode frequencies and $u$ the isothermal
sound speed.  All other symbols have their usual meaning. 
Errors are $1\sigma$ errors as given in the original papers. The
references are: (1) Grevesse \& Noels (1993); (2) Dziembowski et
al.~(1994); (3) Basu et al.~(1996); (4) Degl'Innocenti et al.~(1997);
(5) Christensen-Dalsgaard et al.~(1991); (6) Libbrecht et al.~(1990);
(7) Elsworth et al.~(1994)  }
\protect\label{solval}
\begin{flushleft}
\begin{tabular}{lrc}
\noalign{\smallskip}\hline\noalign{\smallskip}
quantity & value & source \\
\noalign{\smallskip}\hline\noalign{\smallskip}
$T_{{\rm eff},\odot}$ & $5777 \pm 2.5$ K  \\
$L_\odot$ & $(3.844\pm 0.01)\cdot 10^{33}$ erg/s \\
$R_\odot$ & $(6.959\pm 0.001)\cdot 10^{11}$ cm \\
$(Z/X)_\odot$ & $0.0245\pm 0.001$ & (1) \\
$Y_\odot$ & $0.244$ & (2) \\
          & $0.246$ & (3) \\
          & $0.238 - 0.259$ & (4) \\
$R_{\rm cz}$ & $0.713\pm 0.003\, R_\odot$ & (5) \\
$u$ & & (2) \\
$\nu$ & & (6) \\
      & & (7) \\
\noalign{\smallskip}\hline\noalign{\smallskip}
$t_\odot$ & $(4.57\pm0.02)\cdot 10^9$ yrs & \\
\noalign{\smallskip}\hline\noalign{\smallskip}
\end{tabular}
\end{flushleft}
\end{table}

\section{Solar age determinations}

\subsection{Solutions of the 3-parameter problem}

The usual approach to the solar model problem is to fit simultaneously
luminosity and radius (or, equivalently, effective temperature) with
a model having the solar age. In the models, 
the initial helium content and $\alpha_{\rm MLT}$ are the two parameters
that can be varied and therefore the system has a well-determined 
solution. In calculations without metal diffusion, the metalli\-city $Z$
is no independent parameter, because it is determined by $Y_i$ and the
requirement that $(Z/X)_\odot$ has to be matched by the solar model.
In calculations taking diffusion into account, $Z_i$ is an additional
parameter and $(Z/X)_\odot$ another solar quantity to be
matched. The variations in $Z$, however, are only of order 1\% or smaller
and not important for our results. We therefore do not count $Z_i$ as
a further parameter, although in practice it has been varied to match
$(Z/X)_\odot$ in the calculations presented in this section.

Accordingly, in this section we solve the following 3-parameter
problem: in addition to initial helium content and $\alpha_{\rm MLT}$
(resp.\ $\alpha_{\rm int}$) we want to solve for $t_\odot$. Therefore,
we have to add an additional quantity to be matched. For this, we
chose either the surface helium content $Y_\odot$ or the depth of the
convective envelope $R_{\rm cz}$ (Tab.~\ref{solval}). For both
quantities we have selected three values repesenting the mean value
and the allowed lower and upper limits. 

\begin{table}
\caption{Results of the calculations matching the solar luminosity,
effective temperature and a third quantity (second
column) -- either $Y_\odot$ or $R_{\rm cz}/R_\odot$. The third column
contains the corresponding other quantity for comparison ($Y_{\rm s}$
is the surface He content of the final model). 
Calculations with and without hydrogen/helium-diffusion
(forth column) also differ w.r.t.\ opacities and EOS. Standard
Eddington grey atmospheres were used. $t$ is the age in $10^9$ yrs as
determined.} 
\protect\label{threep}
\begin{flushleft}
\begin{tabular}{l|ccc|ccc}
\hline
Case & $3^{\rm rd}$ par. &  & Diff. &
$\alpha_{\rm MLT}$ & $Y_i$ & $t$\\
\hline
A & $Y_\odot$ & $R_{\rm cz}/R_\odot$ & &\\
\hline
A.1 & 0.238 & 0.693 & n & 2.00 & 0.238 & 9.00 \\
A.2 & 0.244 & 0.700 & n & 1.90 & 0.244 & 8.09 \\
A.3 & 0.259 & 0.720 & n & 1.71 & 0.259 & 6.16 \\
A.4 & 0.238 & 0.722 & y & 1.72 & 0.267 & 4.34 \\
A.5 & 0.244 & 0.727 & y & 1.68 & 0.269 & 4.08 \\
A.6 & 0.259 & 0.743 & y & 1.53 & 0.282 & 2.64 \\
\hline
B & $R_{\rm cz}/R_\odot$ & $Y_{\rm s}$ & & \\
\hline
B.1 & 0.710 & 0.253 & n & 1.78 & 0.253 & 6.86 \\
B.2 & 0.713 & 0.254 & n & 1.77 & 0.254 & 6.71 \\
B.3 & 0.716 & 0.257 & n & 1.73 & 0.257 & 6.38 \\
B.4 & 0.710 & 0.233 & y & 1.80 & 0.260 & 5.15 \\
B.5 & 0.713 & 0.235 & y & 1.78 & 0.263 & 4.81 \\
B.6 & 0.716 & 0.240 & y & 1.73 & 0.265 & 4.48 \\
\hline
\end{tabular}
\end{flushleft}
\end{table}

Table~\ref{threep} summarizes the computational details and the
resulting solar ages. For the calculations without
hydrogen/helium-diffusion we used the Saha-type equation of state
(including partial degeneracy) and the Rogers \& Iglesias (1992)
opacities combined with those of Weiss et al.\ (1990) for the low
temperatures. The calculations with diffusion made use of the Iglesias
\& Rogers (1996) plus Alexander \& Fergusson (1994) opacities and the
OPAL-EOS. For all cases listed in Tab.~\ref{threep} standard Eddington
grey atmospheres were assumed and calculations began on the zero-age
main-sequence.

If the helioseismological helium abundance of the photosphere is to be
matched (case A), the neglect of diffusion leads to much too high ages
(up to 97\%) due to the low initial helium value, which results in
luminosities too low that have to be compensated by a higher age. With
diffusion, the ages deviate between 5 and 42\%. Interestingly, case
A.4, with the lowest $Y_\odot$ possible is the model providing the
best fit,
indicating that diffusion might be somewhat too efficient in our
calculations. This agrees with results by Richard et al.\ (1997), who
found that rotationally induced mixing below the convective zone,
counteracting diffusion, not only reproduces lithium and beryllium
depletion correctly, but also improves the agreement in sound speed. 

If the depth of the convective zone is to be matched, again the models
without diffusion tend to have ages too high. (Note that model B.3 is
almost equivalent to A.3.) Models including diffusion
determine a solar age that is within 13\% of the true value. The age
of model B.6, which stretches the uncertainties in  
$R_{\rm cz}/R_\odot$ and $Y_\odot$, deviates by only 2\%. 

It has to be recalled that the models do not include the same input
physics as our solar models (Schlattl et al.\ 1997) in all
respects. However, those including diffusion differ only in the
treatment of the atmosphere and the outer convective regions. They
provide more correct ages, but still deviate by up to
30\%. Therefore, we can conclude that in the 3-parameter case, given
the observational uncertainty in the third parameter, the solar age
cannot be determined as accurate as 10\% for most choices. 
The exceptions are cases B.4, B.5 and B.6, where $R_{\rm cz}/R_\odot$ is
matched but $Y_\odot$ ignored.

\subsection{Fitting 4 quantities with 3 parameters}

In the previous section it became obvious that the depth of the
solar convective zone and the surface helium content cannot be fitted
simultaneously by one model even when using present state-of-the-art
input physics. Consequently, we are now going to
determine the minimum $\chi^2$-fit to all four solar quanitites.
$\chi^2$ is defined as
\begin{equation}
\chi^2 = \sum_{i=1}^{4} {W_i} = \sum_{i=1}^{4} {(A_i - B_i)^2\over
\sigma_i^2}
\label{e:chi}
\end{equation}
where $A_i$ resp.\ $B_i$ denote the observed and model value of
effective temperature, luminosity, surface helium content and depth of
convective zone, and $\sigma_i$ the observational errors (see
Tab.~\ref{solval}). 

\begin{figure}
\epsfxsize=0.85\hsize\epsfbox{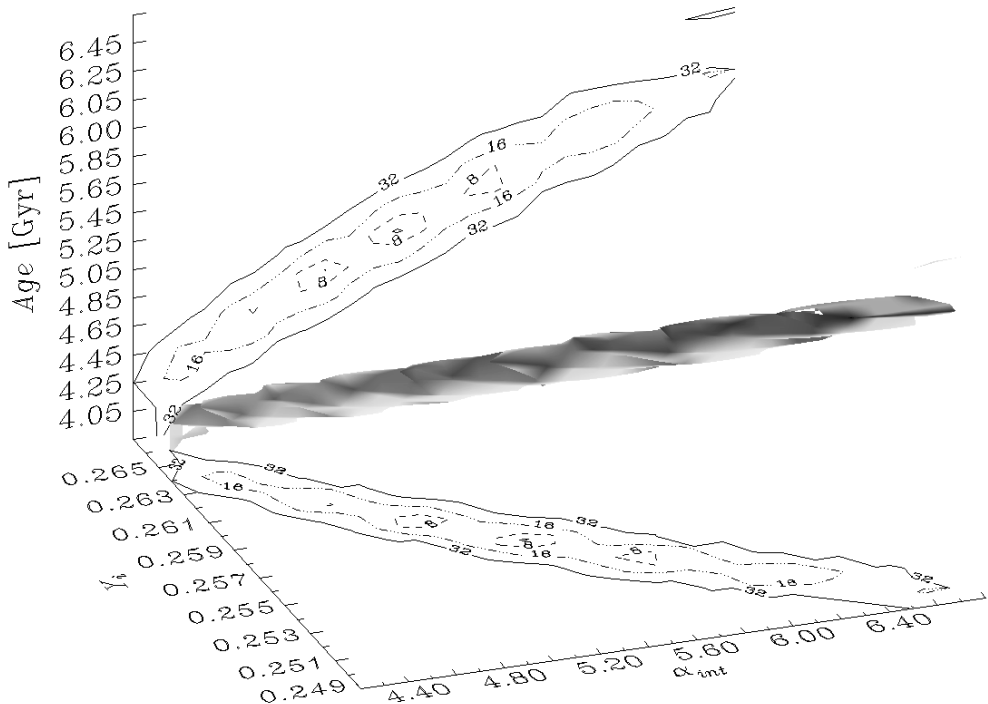}
\caption{Fitting four observational
quantities: $\chi^2=32$ surface and projections onto the $t-\alpha_{\rm
int}$ and $Y_i-\alpha_{\rm int}$ planes in the parameter space }
\protect\label{chiff}
\end{figure}

The models for this set of calculations include, with the exception of
metal diffusion, all the physics of the
best models of Schlattl et al.\ (1997), i.e. the special treatment of
the convective gradient (Eq.(\ref{e:alpha})) and the inclusion of model
atmospheres. They also start with the pre-main sequence
evolution. Otherwise they agree with those of set B.

Contrary to the calculations of the previous section, for the present
problem we have to scan the 3-dimensional parameter space. We did this
with the following steps: $\triangle Y_i = 0.001$, $\triangle
\alpha_{\rm int} = 0.1$, $\triangle t = 2\cdot 10^8$ yrs.
For $\alpha_{\rm int} = 5.5$, we also used a finer $t$-grid of $5\cdot
10^7$ yrs (see below).
As discussed in the previous section we now
kept $Z_i$ constant at the value needed for $(Z/X)_i=(Z/X)_\odot$ in
order to save computing time. In the investigation by Richard et al.\ (1997) 
the influence of varying $Z$ on $Y_i$ and the resulting solar model
can be seen to be very small. We therefore do not expect that our
results would change if we had not made this simplification.

\begin{table*}
\caption{Models for selected values of $\alpha_{\rm
int}$ with the lowest $\chi^2$  of the problem fitting four observational
quantities. $\alpha_{\rm int}$ is the interior value of the variable
mixing-length parameter (see eq.~1). For more details of the models,
see the text. Ages are in $10^9$ yrs.} 
\protect\label{chift}
\begin{flushleft}
\begin{tabular}{l|lcc|cccc|r}
\hline
Case & $\alpha_{\rm int}$ & $Y_i$ & $t$ & $ L/L_\odot$ & $ T_{\rm
eff}$ [K] & $R_{\rm cz}/R_\odot$ & $Y_s$ & $\chi^2$ \\
\hline
C.1 & 4.5 & 0.263 & 4.45 & 1.0051 & 5780.0 & 0.7211 & 0.2359 & 14.84\\
C.2 & 4.7 & 0.261 & 4.65 & 1.0030 & 5777.0 & 0.7186 & 0.2336 &  7.36\\
C.3 & 5.0 & 0.259 & 4.85 & 1.0010 & 5778.0 & 0.7157 & 0.2316 &  3.91\\
C.4 & 5.1 & 0.259 & 4.85 & 1.0014 & 5782.4 & 0.7152 & 0.2317 &  8.27\\
C.5 & 5.3 & 0.257 & 5.05 & 0.9988 & 5777.1 & 0.7138 & 0.2294 &  3.46\\
C.6 & 5.5 & 0.255 & 5.25 & 0.9959 & 5771.3 & 0.7113 & 0.2271 & 11.58\\
C.6'& 5.5 & 0.256 & 5.15 & 0.9977 & 5771.2 & 0.7119 & 0.2284 &  4.51\\
C.7 & 5.7 & 0.255 & 5.25 & 0.9965 & 5778.7 & 0.7108 & 0.2273 &  6.41\\
C.8 & 5.9 & 0.254 & 5.45 & 1.0035 & 5781.9 & 0.7097 & 0.2258 & 12.18\\
C.9 & 6.1 & 0.252 & 5.65 & 1.0001 & 5774.5 & 0.7070 & 0.2235 & 10.73\\
\hline
\end{tabular}
\end{flushleft}
\end{table*}

Figure~\ref{chiff} shows $\chi^2$-surfaces and projected contour lines
for our models. Apparently, there is no clearly defined sharp minimum
which would allow a unique determination of the solar age from fitting
the four quantities. Rather, there exists a long-stretched region of
equally well fitting sets of parameters (including age). The separated
minima visible in the projected contour plots are an artifact
resulting from the grid in parameter space and the plot
interpolation. In reality, the $\chi^2=8$ contour lines are
connected. We have verified this by using a higher resolution in age
for the case of $\alpha_{\rm int} = 5.5$ (cf. Figs.~\ref{chiaa} and
\ref{chiac}). Our grid of models in the $\alpha_{\rm int}$ and $Y_i$
parameter space is rather coarse due to the otherwise inhibitive
number of solar evolution calculations.  In Tab.~\ref{chift} we list
those models with the lowest $\chi^2$ for several values of
$\alpha_{\rm int}$. None of them is fitting all quantities very well
(in particular, $Y_\odot$), as is indicated by the $\chi^2$-values
which should be close to 1 for a good model. This is a confirmation of
the fact that at the presently given accuracy of observations the
theoretical models still contain significant deficits.  Model C.5 is
the best model of the real sun. Its age is within 10\% of the presumed
solar age. C.3 is within 6\% of the true solar age at a slightly
higher $\chi^2$. Given the uncertainty of $\approx$ 4\% in the age
determinations due to the time resolution we use, the true minimum
could well be within 2-3 \% of the solar age.  The effect of a finer
resolution in $\triangle t$ is demonstrated by model C.6'. Relative to
C.6 the minimum is now located more accurately ($\chi^2=4.52$). The
age of the true minimum is between 5.1 and 5.2 Gyr.  It is evident
that all models in Tab.~\ref{chift} have a surface helium content too
low and consequently a rather low initial one, too.

\subsection{Age determination by best-fit models to all observational
data} 

Up to now we have used only two quantities derived from
helioseismology. We are now going to include results about p-mode
frequencies and sound speed. A simple inclusion of many frequencies or
of the sound speed at many radial points as additional degrees of
freedom would dominate the standard $\chi^2$-formula completely. 
In addition, since neither the frequencies nor the sound speed are
statistically independent from $R_{\rm cz}$ or $Y_\odot$, the strict
interpretation of $\chi^2$ is no 
longer valid. We therefore modify Eq.(\ref{e:chi}) to include two
additional terms and interprete the resulting quantity as a function
of merit, defined as
\begin{equation}
\tilde{\chi}^2 = \sum_{i=1}^6 {W_i},
\label{e:chiw}
\end{equation}
where the first four terms are defined as in Eq.(\ref{e:chi}).
The first additional term
\begin{equation}
W_5 \equiv W_{\rm u} = {\int_0^1 (u_{\rm o}(x)-u_{\rm th}(x))^2 dx\over 
             \int_0^1 \sigma_{u}^2(x) dx}
\label{e:chiwu}
\end{equation}
describes the deviation of the model sound speed $u_{\rm th}$ as a function of
fractional radius $x=r/R_\odot$ from the
helioseismologically inferred one ($u_{\rm o}$) weighted by the
observational error $\sigma_{u}$.
The other term
\begin{equation}
W_6 \equiv W_\nu = {\sum_{n=12}^{26} (\nu_{\rm o}(0,n)-\nu_{\rm th}(0,n))^2 \over 
        \sum_{j} \sigma_{\nu}^2(0,n)}
\label{e:chiwp}
\end{equation}
similarly is used to calculate the model goodness of the 15
frequencies $\nu(0,n)$ for a representative mode. We have selected
the $l=0$ mode because it traces the whole solar interior. The choice
of the 15 radial modes was dictated by the availability in the
original papers.

Since the observational errors on $u$ and $\nu$ are extremely small,
even the 
mean deviations $W_{\rm u}$ and $W_\nu$ still would dominate the
function of merit. To allow 
the global quantities to retain their influence, we normalize the
individual terms in Eq.(\ref{e:chiw}) by their
respective minimum values.

\begin{figure}
\centerline{
\epsfxsize=0.75\hsize\epsfbox{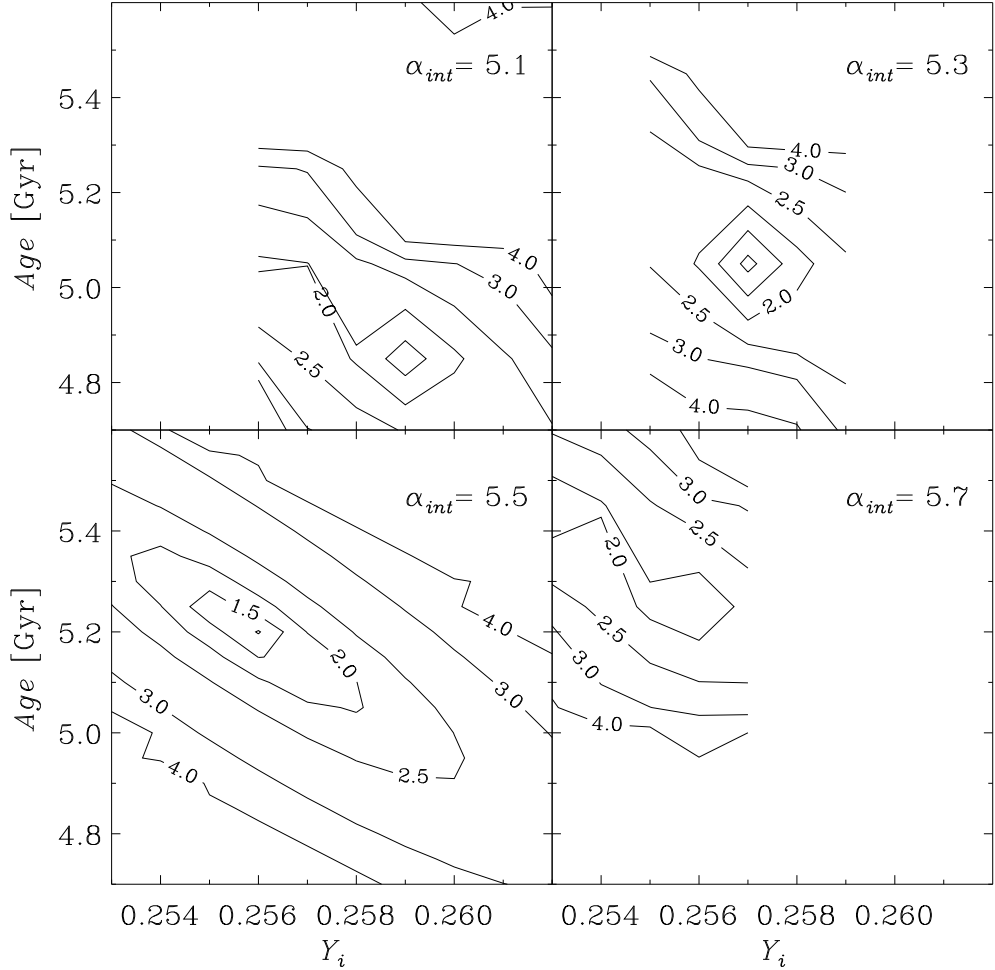}
}
\caption{Quality of model fits using all solar quantities, illustrated
by contour lines of the logarithm of the function-of-merit
$\tilde{\chi}^2$, where all contributing terms have been normalized by
their individual minimum value. The four panels are (from upper left
to lower right) for $\alpha_{\rm int}=5.1,\,5.3,\,5.5,\,5.7$. Only
parts of the parameter space has been investigated}
\protect\label{chiaa}
\end{figure}

\begin{figure}
\centerline{
\epsfxsize=0.7\hsize\epsfbox{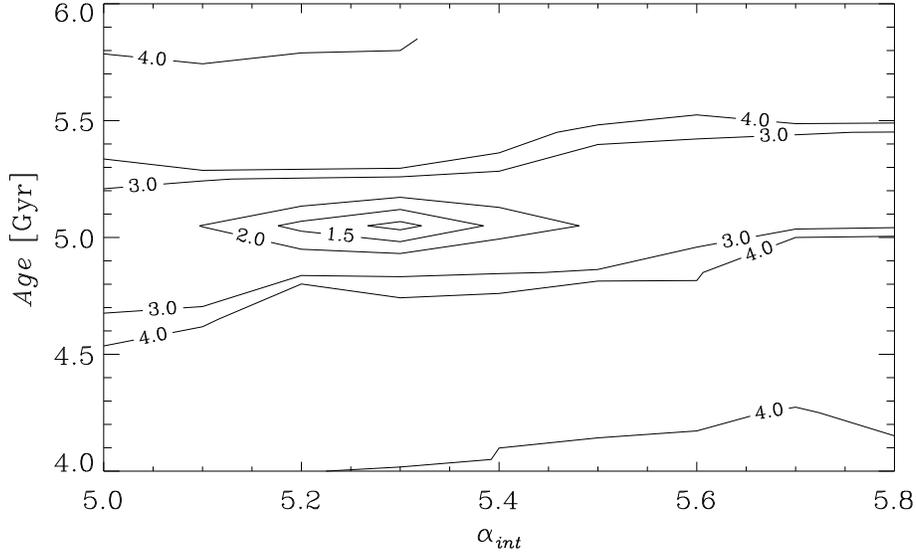}
}
\caption{As Fig.~2, but for the case of fixed $Y_i=0.257$. The
minimum at the center corresponds to model C.5}
\protect\label{chiab}
\end{figure}

\begin{figure}
\centerline{
\epsfxsize=0.7\hsize\epsfbox{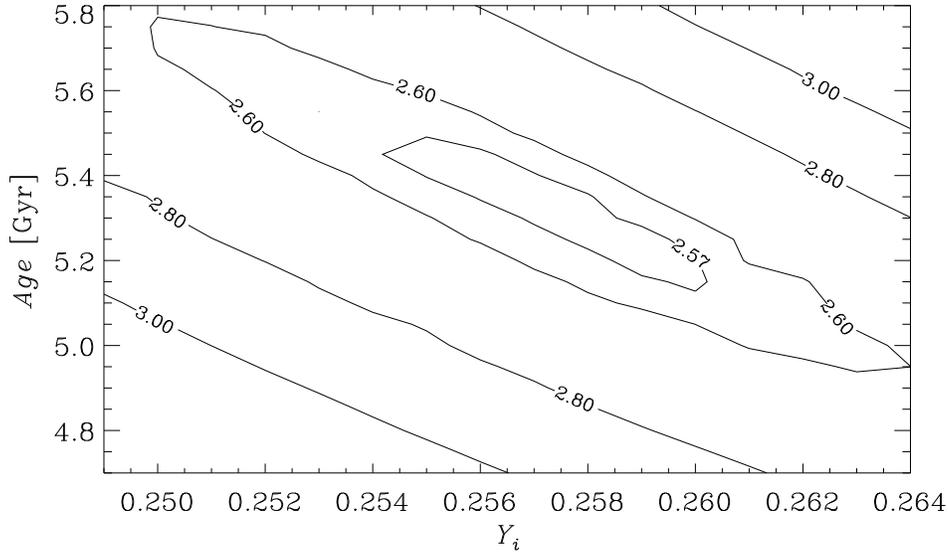}
}
\caption{Contour-lines of $\log W_{\rm u}$ illustrating the quality of
the models with respect to sound speed (fixed $\alpha_{\rm
int}=5.5$; fine $t$-resolution). Note that $W_{\rm u}$ has not been
normalized to its global minimum here}  
\protect\label{chiac}
\end{figure}

For evaluating $\tilde{\chi}^2$ the same solar evolution calculations
as in the last section were used. The result of this age determination
does not differ significantly from that of case C. In Fig.~\ref{chiaa}
we show contour lines of $\log \tilde{\chi}^2$, the function-of-merit
as defined by Eqs.(\ref{e:chiw}-\ref{e:chiwp}), where the individual
terms have been normalized by their respective minimum
value. Evidently, there are only very small regions, where
$\tilde{\chi}^2 < 10$, for all four cases of $\alpha_{\rm int}$
shown. For $\alpha_{\rm int}$ smaller than 4.7 and larger than 6.1
they vanish completely. The 3-dimensional isocontour resembles that of
Fig.~\ref{chiff} very much, therefore.  The effect of a finer
resolution in age ($\triangle t = 5\cdot 10^7$ yrs) is seen in the
lower left panel ($\alpha_{\rm int}=5.5$). Here, the contours for $\log
\chi^2 < 3$ are still elongated, while in all other panels they form
isolated wells. We have compared the shown $\alpha_{\rm int}=5.5$ contour
lines with those for the same $t$-resolution as used for the other three
panels ($\triangle t = 2\cdot 10^8$ yrs) and found the same round
shape of the $\log \chi^2 < 3$ contours. The minimum for this case
deviates slightly from that of model C.6', which was calculated with
the same resolution. It is now at 5.2 Gyr but the same $Y_i = 0.256$.
The global minimum is reached (see
Fig.~\ref{chiaa} upper right panel) for $\alpha_{\rm int}=5.3$ at $Y_i
= 0.257$ and age $t_\odot = 5.05 \cdot 10^9$ as for model C.5. For
comparison, a cut at $Y_i=0.257$ is shown in Fig.~\ref{chiab}. The
$\tilde{\chi}^2$ minimum again corresponds to model C.5. 

\begin{figure}
\centerline{
\epsfxsize=0.65\hsize\epsfbox{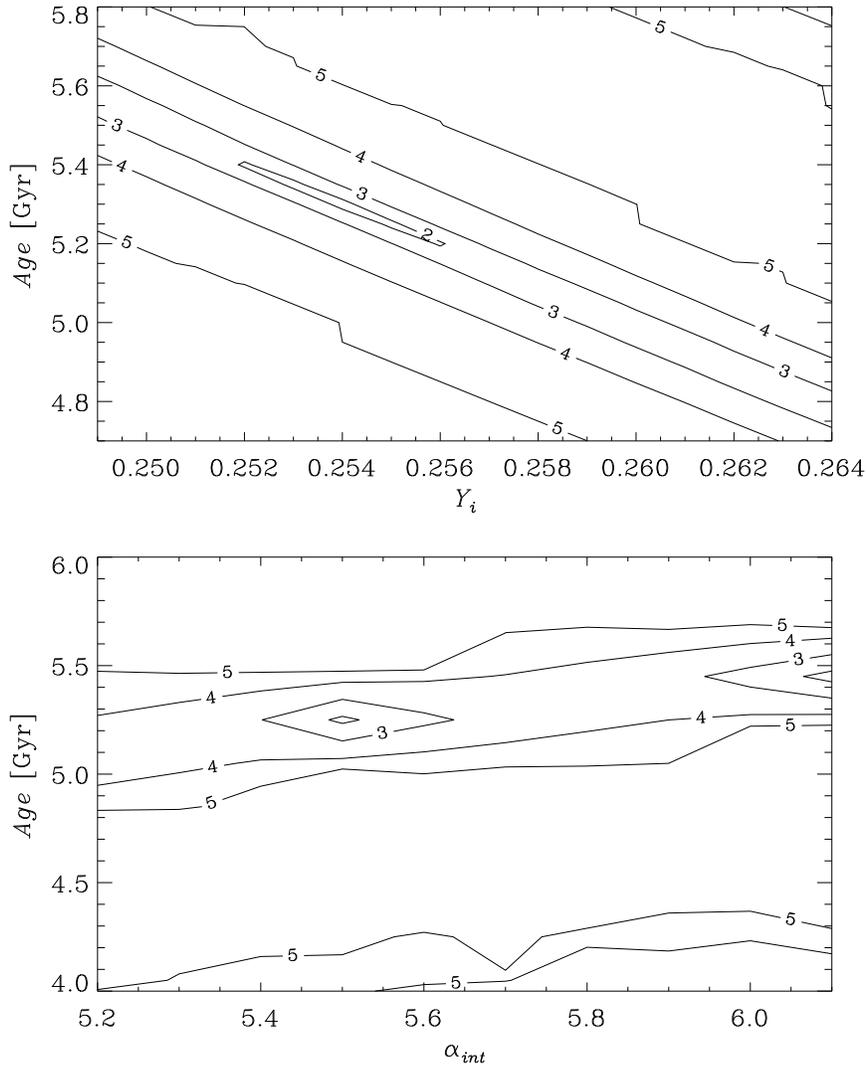}
}
\caption{Contour-lines of the unnormalized $\log W_\nu$
illustrating the quality of the models with respect to the frequencies of
the $l=0$-mode. Upper panel: for fixed parameter $\alpha_{\rm
int}=5.5$ (fine $t$-resolution); lower panel: fixed $Y_i=0.255$}  
\protect\label{chiad}
\end{figure}

The contribution of $W_{\rm u}$ and $W_\nu$ to $\tilde{\chi}^2$ is illustrated in
Figs.~\ref{chiac} and \ref{chiad}. Fig.~\ref{chiac} demonstrates that 
the sound speed $\tilde{\chi}^2$-contours $W_{\rm u}$ outline a long-stretched
valley enclosing the parameters of the best models without defining a
clear minimum. Rather, the function-of-merit appears to be almost
degenerate. Furthermore, the region of smallest $\chi^2$ is shifted
towards higher ages and helium content with respect to the location of
the best model. The shown case ($\alpha_{\rm int}=5.5$) is a typical
one. $u$ cannot be used for age determinations.

Similarly, Fig.~\ref{chiad} shows how the models fit the
$l=0$ p-mode frequencies. The minima, though not very low, are very sharply
defined, and the models providing the best fits for the complete
problem also have a comparably low $W_\nu$. 
Model C.6, e.g., lies within the very narrow $\log W_\nu=2$ contour of
the upper panel, which was plotted using the high $t$-resolution.
The minimum within this countour is close to the
position of model C.6, but at a slightly larger age. 
The position of model C.6', which does not include the fit to the
p-mode frequencies, but was calculated with the same high
$t$-resolution, is outside the $\log W_\nu=2$ contour. Therefore,
at a resolution of $5\cdot 10^7$ years, the minima of models C do no
longer coincide with those of the p-mode fits. The complete fit (see
Fig.~\ref{chiaa}) incidentally is at the position of model C.6. Since
in the complete fit we applied an -- arbitrary -- weighing of the
contributing terms to $\tilde\chi^2$ by the normalization, the
difference between these various minima gives a measure for a
systematic age determination uncertainty of the order of 2\%.

In the lower
panel, Model C.6 can clearly be identified as the central minimum.
The p-modes therefore add very
important information in selecting the best-fit models and determining
the solar age. The $W_\nu$-term also dominates $\tilde{\chi}^2$
and the contours of Fig.~\ref{chiaa}. We cannot exclude that for an
even higher resolution in $\alpha_{\rm int}$ or $Y_i$ the $W_\nu$
mimima deviate from those of the $\tilde\chi^2$  or
of $\chi^2$ of models C, similar as in the demonstrated case of a
higher $t$-resolution.

\clearpage

\section{Discussion}

If the solar age were to be determined from stellar models fitting
observed solar quantities, we have shown that a priori an uncertainty
of a factor of two would exist. The reason for this is the fact that
apart from the global solar values -- radius, effective temperature
and $(Z/X)_\odot$ -- the third quantity necessary to restrict mixing
length parameter, initial helium content and age, has to be one
derived from helioseismology. In our calculations, this was either the
depth of the solar convective zone or the present photospheric helium
abundance. Both quantities depend crucially on the effect of particle
diffusion, though $R_{\rm cz}$ does less so (Tab.~\ref{threep}). 
If this is not recognized, and diffusion not taken into 
account into the calculations, derived ages are by far too high (model
sequences A \& B without diffusion). If diffusion is taken into
account, the predicted ages can be as accurate as 10\%, if $R_{\rm cz}$
is the third parameter, but are too low by up to 40\% for the allowed
range of $Y_\odot$. The latter fact indicates that either diffusion is
too effective in the models or that $Y_\odot$ is rather at the lower
boundary of the present determinations.

Including both of these helioseismological quantities in fitting the
present sun, ${\chi}^2$-fits using ``best-physics-models'' yield ages
that agree to 10\% or better with the solar age. However, none of the
models is a good model for the Sun with ${\chi}^2 \approx 1$ and there
is no clear global ${\chi}^2$-minimum fit. If we generously accept all
models of Tab.~\ref{chift} with ${\chi}^2 \la 10$ (approx.\ $3\sigma$),
the solar age could range between 4.65 and 5.65 Gyr, being slightly
too high. Therefore, a systematic error in the models is still
present.

Adding sound speed and p-mode frequencies to the fits, the
best-fitting models remain nearly the same as in the previous case, if we
normalize the individual contributions to the function of merit. This
is necessary, since all solar models do not predict p-mode frequencies
and sound speed throughout the Sun with sufficient accuracy for the
extremely low errors in the observations. 
Since the various quantities used in this fit
are no longer independent of each other, the function of merit, though
still denoted by $\tilde{\chi}^2$, no longer has any statistical meaning.
While sound speed does not
provide helpful information for age determinations, the p-mode
frequencies are extremely sensitive to deviations of the models from
the observed Sun. In fact, though all models are poor fits for the
p-modes, the function-of-merit minima are extremely sharp and coincide
with the parameters also yielding the best fits for the previous case
of 4 solar parameters, as long as we use age-steps of $2\cdot 10^8$
yrs. At a four times higher resolution, the various
$\tilde\chi^2$-minima do no longer coincide, but the location of the
``true'' minimum remains subject to the exact definition of
$\tilde\chi^2$.  p-mode frequencies alone might suffice for
finding best-fit parameters, including the solar age.

\begin{table*}
\caption{$\chi^2$-minima  of the problem fitting four observational
quantities for $\alpha_{\rm int}=5.3$ and two values of $Y-i$. In
these calculations metal diffusion has been taken into account and the
high resolution in time has been applied. For comparison, model C.5 is
repeated} 
\protect\label{chiftm}
\begin{flushleft}
\begin{tabular}{l|lcc|cccc|r}
\hline
Case & $\alpha_{\rm int}$ & $Y_i$ & $t$ & $ L/L_\odot$ & $ T_{\rm
eff}$ [K] & $R_{\rm cz}/R_\odot$ & $Y_s$ & $\chi^2$ \\
\hline
D.1 & 5.3 & 0.257 & 5.00 & 0.9970 & 5779.8 & 0.7182 & 0.2295 &  8.40\\
D.2 & 5.3 & 0.256 & 5.15 & 0.9994 & 5774.5 & 0.7180 & 0.2277 &  9.42\\
\hline
C.5 & 5.3 & 0.257 & 5.05 & 0.9988 & 5777.1 & 0.7138 & 0.2294 &  3.46\\
\hline
\end{tabular}
\end{flushleft}
\end{table*}

We have for one special case investigated the effect of metal
diffusion ($\alpha_{\rm int} = 5.3$; $\triangle t = 5\cdot 10^7$; cf.\
model C.5). Table~\ref{chiftm} lists the minima for two different
choices of $Y_i$. Compared to C.5, model D.1 (same $Y_i$) has a
minimum of higher $\chi^2$ at a slightly lower age. A better fit is
found for a lower $Y_i$ but at a higher age (D.2).
At least for this value of $\alpha_{\rm int}$, we conclude that the
consideration of metal diffusion would lead to a $\chi^2$-minimum at a
much higher age. Whether the global $\chi^2$-minimum would also be
shifted towards an higher age, can only be decided after the
(time-consuming) calculations for all parameters have been done. From
the fact that the depth of the convective zone has increased for
models D, we infer that only the C-models with a low value for $R_{\rm
cz}/R_\odot$ could give the global $\chi^2$-minimum, if metal diffusion
were included. Those, however, already have ages too high. It might be
that the inclusion of ``better'' physics in this case leads to a
larger error in the age determination.

The implications for stellar age determinations are not
straightforward. Evidently, the physical input used today for the
standard solar model (including diffusion) provides the most
accurate age determination. It therefore should be used in stellar
models in general. If direct measurements of stellar parameters (mass,
luminosity, effective temperature) are available, the derived ages can
be expected to be comparably accurate as in the solar case provided
the observations are as accurate as for the Sun.

The inclusion of diffusion is the most important aspect of solar age
determinations performed as in the present paper. If it is neglected,
the derived ages are unacceptable. This is due to our knowledge about
the helium content or depth of the solar convective zone. For other
stars, this information will not be available directly. Evolutionary
ages will then depend solely on the assumed initial helium content, no
matter whether diffusion is taken into account or not. Paczy\'nski
(1996a) argues that detached eclipsing binaries will provide the
possibility to determine age and (initial) helium content at the same
time. In this case, diffusion most likely has to be included, since
the central helium content is changed by diffusion and this influences
the main-sequence lifetime. Otherwise, an additional uncertainty of
about 5\% (Chaboyer 1995) is added. Note that for old systems, the
relative error due to a wrong helium content is smaller than in the
solar case, because it leads to a luminosity change roughly constant
for the whole main sequence evolution. Absolute errors could
be of up to a few Gyr.

For age determination methods based on differential quantities like
the brightness difference between turn-off and horizontal branch
the situation is more complicated. If the defects of the stellar
evolution calculations, which lead to the solar age errors, affect the
different stellar evolution stages systematically, the ages based on
differential methods could be more accurate than the results found
here indicate. (There are, however, many other, probably more severe
sources of errors, like the conversion from theoretical to
observational brightness.) This, however, has to be investigated in
detail (Castellani et al.\ 1997).

To summarize, we have demonstrated that even with the best
observational data available, present stellar evolution calculations
cannot be expected to yield ages with errors less than
10\%. Hydrogen/helium diffusion is an absolutely necessary ingredient
to reach this level of accuracy. We have not investigated, whether the
inclusion of metal-diffusion would lead to a more accurate solar age,
because in most stellar evolution calculations this most likely
will not be included due to computational limitations.

\begin{acknowledgements}
It is a pleasure to thank B.~Paczy\'nski for initiating the present
work by asking us a simple, but fundamental question and for his
continuing support and stimulating discussions.
H.S.\ was supported by the ``Sonderforschungsbereich 375-95 f\"ur
Astro-Teilchenphysik'' of the Deutsche Forschungsgemeinschaft.
\end{acknowledgements}


\begin{thebibliography}{99}
%
\bibitem{AF.94} Alexander D.R., Fergusson J.W., 1994, ApJ 437, 879
\bibitem{BCSTT.96} Basu S., Christensen-Dalsgaard J., Schou J.,
Thompson M.J., Tomczyk S., 1996, ApJ 460, 1064
\bibitem{Cast.96} Castellani V. et~al., 1994, Phys.~Rev.~C 50, 4749
\bibitem{Cast.97} Castellani V., Ciacio F., Degl'Innocenti S.,
Fiorentini G., 1997, preprint astro-ph~9705035
\bibitem{Chab.95} Chaboyer B., 1995, ApJL 444, 9
\bibitem{ChK.95} Chaboyer B., Kim Y.-C., 1995, ApJ 454, 767
\bibitem{ChGT.91} Christensen-Dalsgaard J., Gough D.O., Thompson M.J.,
1991, ApJ 378, 413
\bibitem{DCM.97} D'Antona F., Caloi V., Mazzitelli I., 1997, ApJ
477, 519
\bibitem{Deg.97} Degl'Innocenti S., Dziembowski W.A., Fiorentini G.,
Ricci B., 1997, preprint astro-ph~9612053
\bibitem{DGPS.94} Dziembowski W.A., Goode P.R., Pamyatnych A.A.,
Sienkiewiecz R., 1994, ApJ 432, 417
\bibitem{Els.94} Elsworth Y. et~al., 1994, ApJ 434, 801
\bibitem{GN.93} Grevesse N., Noels A., 1993, Phys.~Scripta T47, 133
\bibitem{IR.96} Iglesias C.A., Rogers F.J., 1996, ApJ 464, 943
\bibitem{KKS.96} Kaluzny J., Kubiak M., Szyman\'nski M., Udalski
A., Krzemin\'ski W., Mateo M., 1996, A\&AS 120, 139
\bibitem{LWK.90} Libbrecht K.G., Woodard M.F., Kaufman J.M., 1990, ApJS 74, 1129
\bibitem{MDC.95} Mazzitelli I., D'Antona F., Caloi V., 1995, A\&A
302, 382 
\bibitem{Pac.96a} Paczy\'nski B., 1996, in {\bf xxx} (ed.), Variable  
stars and the astrophysical returns of microlensing surveys, IAP, 
July 7-12, 1996, in press   
\bibitem{Pac.96b} Paczy\'nski B., 1996, in Turok N. (ed.), Critical
Dialogues in Cosmology, Proceedings of the Princeton University,
Conference June 24-27, 1996, Princeton, University Press, in press
\bibitem{RVCD.97} Richard O., Vauclair S., Charbonnel C., Dziembowski
W.A., 1997, A\&A, submitted
\bibitem{RI.92} Rogers F.J., Iglesias C.A., 1992, ApJS 79, 507
\bibitem{RSI.96} Rogers F.J., Swenson F.J., Iglesias C.A., 1996, ApJ 456, 902
\bibitem{SDW.97} Salaris M., Degl'Innocenti S., Weiss A., 1997, ApJ
479, 665
\bibitem{SchW.97} Schlattl H., Weiss A., Ludwig H.-G., 1997, A\&A, in press
\bibitem{SPE.97} Schr\"oder P., Pols O.R., Eggleton P., 1997, MNRAS
285, 696
\bibitem{Shi.95} Shi X., 1995, ApJ 446, 637
\bibitem{SVB.96} Stetson P.B., VandenBerg D.A., Bolte M., 1996, PASP
108, 560 
\bibitem{TBL.94} Thoul A.A., Bahcall J.N., Loeb A., 1994, ApJ 421, 828
\bibitem{Wei.97} Weiss A., Schlattl, H., 1997, in preparation
\bibitem{WKM.90} Weiss A., Keady J.J., Magee N.H., 1990, Atomic Data
and Nuclear Data Tables 45, 209 
%
\end{thebibliography}
\end{document}